# Modeling Biological Membrane and Red Blood Cells by Coarse-Grained Particle Method

He Li[1*], Hung-Yu Chang[1], Jun Yang[2], Lu Lu[1], George Lykotrafitis[3]

[1]Division of Applied Mathematics, Brown University, Providence, Rhode Island, 02912, U.S.A

[2]Department of Materials Science and Engineering, Massachusetts Institute of Technology, Cambridge, Massachusetts, 02139, U.S.A.

[3]Department of Mechanical Engineering, University of Connecticut, Storrs, Connecticut, 06269, USA

**ABSTRACT**

In this work, we review previously developed coarse-grained (CG) particle models for biological membrane and red blood cells (RBCs) and discuss the advantages of the CG particle method over the continuum and atomic simulations on modeling biological phenomena. CG particle models can largely increase the length scale and time scale of atomic simulations by eliminating fast degrees of freedom while preserving the mesoscopic structures and properties of the simulated system. On the other hand, CG particle models can be used to capture microstructural alternations in diseased RBCs and simulate topological changes of biological membrane and RBCs, which are major challenges to typical continuum representations of membrane and RBCs. The power and versatility of the CG particle methods are demonstrated through simulating the dynamical processes involving significant topological changes, such as lipid self-assembly, vesicle fusion and membrane budding.

*Corresponding author He Li, Ph.D E-mail: he_li@brown.edu

# 1. Introduction

Biological membrane is a vital component of living cells as it helps preserve the cell integrity[1, 2]. Biomembrane also plays a key role on various functions of cells [3]. Lipid vesicles formed by budding off from a cell membrane can transport biological materials between cells. These transport vesicles drop off their loads through fusion with the membrane of the target cells. In addition, cell membrane is also essential in multiple intercellular processes such as cell differentiation, cell–cell adhesion, and cell migration [1]. Biological membrane consists of various lipid molecules and proteins. A typical lipid molecule is constituted by two hydrophobic hydrocarbon chains and a hydrophilic polar head, such that lipid molecules can spontaneously aggregate into bilayer and then form vesicles in aqueous environment because of the hydrophobic effect. Biomembrane behaves similar to two-dimensional (2D) fluid. Despite that the thickness of membrane is only ~ 5 nm, the size of lipid vesicles can be up to micrometers [2].

Red blood cells (RBCs) are unique amongst eukaryotic cells as they bear no nucleus or cytoplasmic structures or organelles [4]. Therefore, the structural properties are linked to the cell membrane. In addition to the lipid bilayer and integral proteins, RBC membrane possesses a 2D cytoskeleton tethered to the lipid bilayer. The RBC membrane cytoskeleton has a 2D six-fold structure and it is made of spectrin tetramers, connected at the actin junctional complexes, forming a 2D six-fold structure. The cytoskeleton is connected to the lipid bilayer via ''immobile'' band-3 proteins at the spectrin-ankyrin binding sites and via glycophorin protein at the actin junctional complexes [4]. In RBC membrane, the stiffness and elasticity of the RBCs arise primarily from the cytoskeleton. Given this particular membrane structure, RBCs have remarkable deformability which allows them to undergo repeated severe deformation when transversing through small blood vessels and organs [5].

In spite of the significant advances made in computing power in the past few decades, it is still computationally prohibitive or impractical to perform atomistic simulations on cell membrane at time scales and length scales that can be used to directly compare with typical laboratory experimental studies. The atomistic simulation techniques are limited by the number of atoms/molecules involved, typically $10^4$~$10^8$ corresponding to a length scale on the order of tens of nanometers. On the other hand, the maximum time step in atomistic simulations is limited by the smallest oscillation period of the fastest atomic motions in a molecule, which is typically several femto-seconds ($10^{-15}$s). Most atomistic membrane models have been limited to the study of only a few hundred lipids or single proteins for a period of a few nanoseconds because of the excessive computational cost [6-11].

At the opposite end of the length scale and time scale spectrums, continuum models [12-14] were developed to simulate morphologies of vesicle and to measure thermal fluctuations of fluid membranes at much larger length scales [15, 16]. Boundary integral method [17-20] and immersed boundary method [21-24] are two popular ways to simulate RBCs based on the assumptions that RBC membrane and embedding fluids are homogeneous materials. These continuum-based RBC models facilitate the study of single cell dynamics, such as tank-threading [25], RBC passing splenic slit [26], or blood flow on macroscopic length and time scales [27, 28]. Although the continuum-based RBC models provide an accurate description of RBC deformation at cellular level, they are not able to describe mesoscopic- and microscopic-scale phenomena, such as representing the structural defects in the RBC cytoskeleton, or capture the topological changes of membrane or RBCs, such as self-assembling, fusion of lipid vesicles and membrane budding.

The limitations of atomistic and continuum methods have motivated a continuous effort on developing coarse-grained (CG) particle models that bridge atomistic and continuum models [10, 29-51]. CG particle models can substantially simplify the atomistic dynamics by eliminating fast degrees of freedom while preserving the mesoscopic structures and properties of the simulated system [52] and it has been applied broadly in polymer dynamics [53-59] and biological fibers [60-66]. In the following sections, we will review the previously developed biological membrane models, RBC membrane models and RBC whole cell models and demonstrate their power and versatility in simulating the lipid self-assembly, vesicle fusion and membrane budding.

## 2. Coarse-grained particle model
### 2.1 Biological membrane models

CG membrane models treat a group of atoms in a single phospholipid molecule into CG particles as phospholipids are the most abundant membrane lipids which have a polar head group and two hydrophobic hydrocarbon tails. An atomic model for a phospholipid molecule is shown in Fig. 1(a). As a result, each lipid molecule is modeled as a short chain of CG particles. The levels of coarse-graining in different CG membrane models are determined by the targeted problems of the simulation. Markvoort et al.[45] introduced a coarse-grained molecular dynamics (CGMD) model for lipid molecules where two chains of four CG particles represent the two hydrophobic tails and another chain of four CG particles represents the lipid headgroup, as shown in Fig.1 (b). The authors used two types of lipids to introduce asymmetry in the lipid bilayers. At a higher level of coarse-graining, Wang and Frenkel [43] proposed a membrane model where each lipid molecule was represented by only three CG particles with one particle being hydrophilic head and the other two being hydrophobic tails (see Fig.1 (c)). A number of CGMD membrane models were developed following similar strategy [30, 31, 33-35, 38, 39, 44, 67]. To achieve a high level of coarse-graining, Yuan et al. [68] coarse-grained a group of lipid molecules to one single CG particle and built a one-particle thick lipid bilayer model (see Fig.1 (d)). The hydrophilic and hydrophobic properties of the lipid molecules are represented by a directional vector.

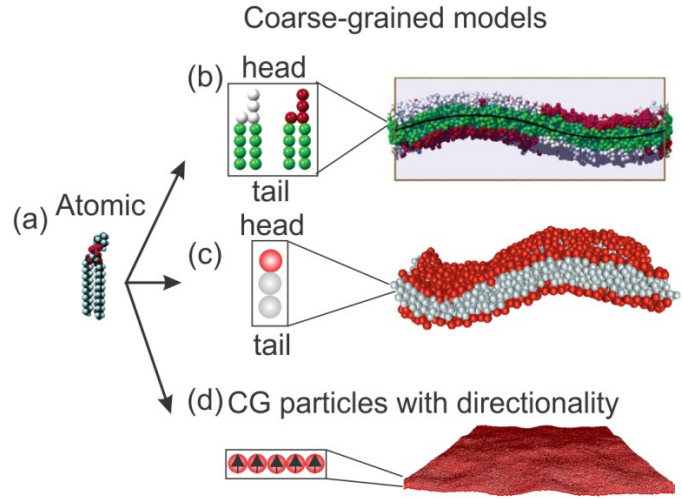

Figure 1. (a) An atomic model for a lipid molecule. (b) A CGMD model representing a lipid molecule by three chains of particles. Two four-particle chains represent the tails of lipid molecule and one four-particle chain represents the head. Adapted from Ref [45]. (c) A CGMD model representing a lipid molecule by a chain of three CG particles. Two CG particles represent the tails of lipid molecule and one CG particle represents the head. Adapted from Ref [43] (d) One-particle thick lipid bilayer model where each CG particle carries a directional vector and it represents a lump of lipid molecules[68].

CG particle models for biological membrane can also be grouped into explicit solvent models and implicit solvent models[29, 44, 48]. In explicit solvent models, employ the hydrophobic interactions are implemented between lipids and solvent particles to maintain the stability of the lipid membrane [34, 42]. Explicit solvent models frequently employ dissipative particle dynamics (DPD), a very efficient method that represents a large volume of the solvents with a soft bead, thus significantly accelerating the computations [69-71]. Three types of forces are applied in traditional DPD models, namely, conserved soft repulsion force, pair-wise dissipation forces and pair-wise random forces. The temperature of the DPD system is preserved through balancing of dissipation and random forces. The momentum of DPD particles is also preserved and thus provides accurate hydrodynamics. Besides increasing the length scale, DPD simulations can apply a larger timestep because of the soft repulsion conservative forces. However, explicit solvent methods can dramatically boost the computational expense as the solvent particles are required to fill a 3D simulation box. The number of solvent particles can grow quickly when size of the simulation box is increased. On the other hand, implicit solvent models can overcome these disadvantages and thus can achieve larger length and time scales [44, 72]. In implicit solvent models, instead of modeling the solvent particles, their effect is represented by effective multi-body interaction potentials, based on either the local particle density [30, 33, 43], or by applying different pair-potentials between particles representing the hydrophobic tails and particles representing the hydrophilic head of the lipids [35, 38, 39]. Development of solvent-free CGMD lipid membrane models requires representations of hydrophobic interactions by interacting potentials so that the hydrophobic behavior of the lipid molecules can be captured. One group of models applied a pair potential between tail beads softer than the LJ-type potentials to preserve the stability of the fluid membrane and then implemented a LJ-type potentials for all the other inter-bead interactions [39]. Along this line, Drouffe et al. [30] and Noguchi et al. [46] developed one-particle-thick, CGMD models to simulate biological membranes. In Drouffe's model, a LJ-type pair potential is applied to simulate the interactions between CG particles. This potential is a function of the distance between the particles and their directionality. Noguchi et al. introduced a multi-body potential that did not require the directionality of particles [46]. Yuan et al. introduced membrane model with application of a soft-core potential in order to illustrate the particle self-diffusion [68].

**2.2 Lipid self-assembling and membrane fusion**
Self-assembly processes of lipids from a randomly dispersed state to an ordered bilayer structure have been captured by CG particle models [36, 45, 73, 74]. Starting from a random dispersion, lipids rapidly aggregate into micelles and small bilayers also referred to bicelles. Subsequently, these small aggregates merge into a large disk-like bilayer. In order to minimize the line tension arising at the bilayer edge, this bilayer tends to seal and form a vesicle by gradually encapsulating water, see a self-assembly process of lipids in Fig. 2. In fact, such bent lamellar structure is essential to cell membrane related processes, especially during exo- and endocytosis. For entropic reasons, a bilayer does not stay at a simply planar state, it exhibits thermal fluctuations (undulations), and its bending rigidity is responsible for manipulating the fluctuations in the average position of the bilayer surface [75]. Although a vesicle can persist for a long time, it is merely in a metastable state instead of a thermodynamic equilibrium state. This is attributed to the implicitly higher pressure and chemical potential of water in the interior domain than those in the exterior domain of a vesicle [74]. In addition, the smaller the vesicle is the larger the pressure difference exists, and thus, small vesicles are thermodynamically less stable against fusion than the larger ones.

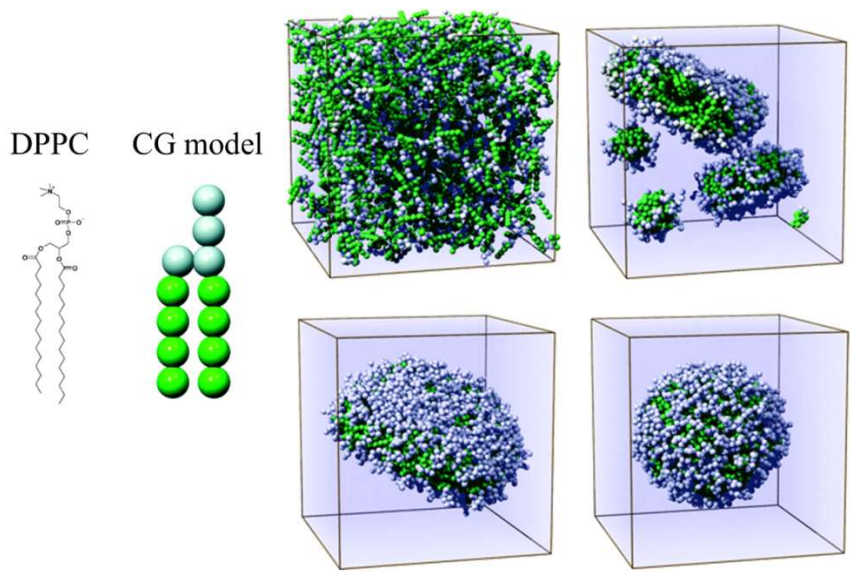

Figure 2. A randomly dispersed DPPC (dipalmitoylphosphatidylcholine) lipids spontaneously form an enclosed bilayer structure (a vesicle), and the snapshots of the evolutionary process are from the work by Markvoort et al. [42].

Lipids in bilayer membrane can be either a solid-ordered state or a liquid-disordered state. It is known that lipid bilayers exhibit a phase transition from a gel phase to a liquid phase as the temperature increases [76, 77]. For some lipids, e.g. DPPC and DMPC (dimyristoylphosphatidylcholine), there could exist an intermediate state (ripple phase) between the gel and liquid phases [78], see Fig 3, and the ripple phase can be either symmetric or asymmetric, depending on the cooling rate from the liquid state down to a crystalline structure [79]. Different phase behavior of a lipid bilayer affects its membrane structure and mechanical properties, and therefore the functioning of biological membranes is also temperature-dependent. In addition to the change of molecular volume (or area per lipid) and heat capacity with temperature as shown in Fig 3, several DPD simulations [80-82] have demonstrated that the membrane thickness of the lipid bilayer ($h$) and the orientation of lipid tail ($S$) can be a function of the temperature ($T$) and the main transition temperature ($T_m$) is located right at the inflection point of $h(T)$ and $S(T)$. CG membrane models can facilitate the estimation of membrane physical properties as well. For example, the water permeability of the modeled membrane was predicted by the release rate of the lumenal substance. Wu et al. [81, 83] have showed that the water permeability of the bilayer membrane grows exponentially with increasing $T$ for $T > T_m$ and their simulation results agreed well with the experimental measurements [83]. In addition, fluctuation spectrum of lipid membrane can provide information of the bending modulus. The intensity $I = \langle u^2 \rangle$ of the undulations follows $q^4$ behavior in the long wavelength regime, i.e. $I \propto k_B T (A k_c q^4)^{-1}$, where $u$ is the fluctuation amplitude, $A$ is the membrane area, $q$ is wavenumber, and $k_c$ is the bending modulus. By fitting the equation to the simulation data, Marrink et al. [84] obtained $k_c = 4 \pm 2 \times 10^{-20}$ J, which is consistent with the experimentally measured $5.6 \pm 0.6 \times 10^{-20}$ J for DMPC lipids [85].

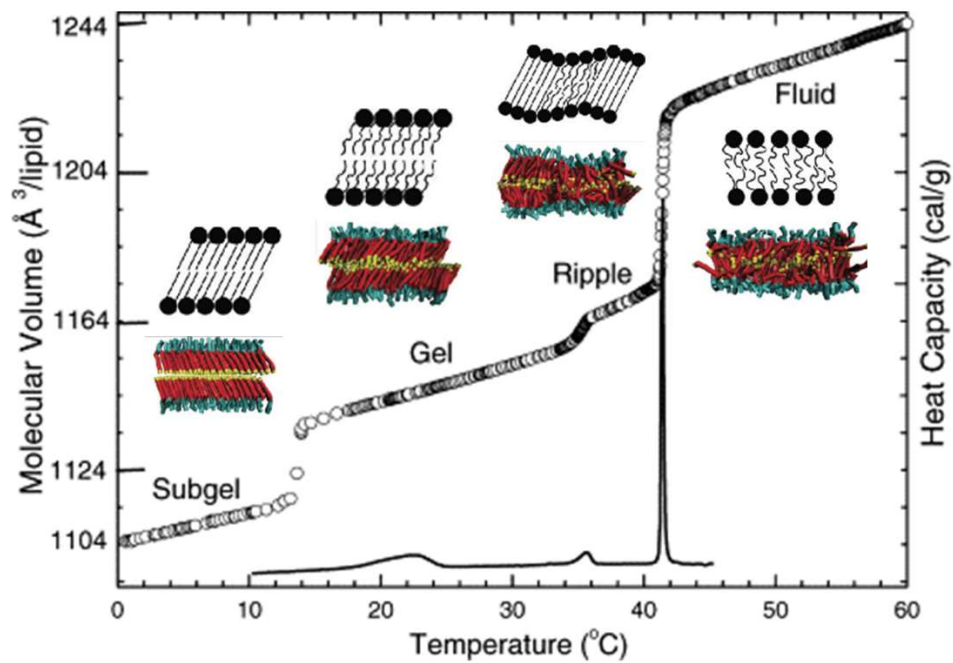

Figure 3. Phase transition of DPPC bilayers in excess water with experimental data of molecular volume (open circles) and heat capacity (solid line) showing a sharp transition from ripple to fluid (liquid) phases at $T_m \sim 41$ °C. Data is adopted from Refs. [78, 86, 87]. Representative diagrams of various lipid phases are from the DPD work of Rodgers et al. [80], and the modeled lipids are colored in cyan for lipid heads, red for lipid tails, and yellow for the final bead in each tail in order to better display the relative tail order in each phase.

Vesicle fusion is a principal function of cells to communicate and execute a series of biophysical activities from vesicular trafficking to cell-cell fusion [88]. In addition, lipids fusion is utilized to several technological applications such as drug delivery [89] and the surface modifications of supported lipid bilayer for biosensors [90, 91]. Many in vitro fusion experiments have been performed with lipid vesicles and the dynamic process is in milliseconds and micrometer length scale [92, 93]. However, the molecular structures and transitions during fusion cannot be captured and visualized by currently experimental techniques. On the other hand, numerical framework can allow itself to elucidate the vesicle fusion in more detail. Several pathways have been proposed to describe the fusion processes of lipid vesicles with the help of CG modeling, and the stalk–pore hypothesis is widely used [37, 94-98], see also Fig. 4 (left). First emergence a neck–like structure (usually regarded as a stalk) was initiated by splayed and tilted lipids that connected the outer monolayers of two approaching vesicles. Then, the stalk grows quickly to form a hemifusion diaphragm (HD), which has thickness equal to a bilayer thickness. The lumenal contents enclosed by each vesicles still remain separated in the HD stage. Finally, the pores are generated to end up the fusion process. By using CG lipid model, Marrink et al. [37] found that alternations of lipid composition can result in distinct speeds for stalk formation and the opening of the fusion pore. Another pathway for lipid fusion is called the direct stalk–pore model [97-101], as shown in Fig. 4 (right). Similar to the original stalk model, the modified one is defined when there is no evident HD stage during the fusion. It occurs usually when vesicles are under high membrane tension and the pores are initiated from the radially expanding stalk. Both the original and modified stalk models are applicable to the fusion of vesicles formed by phospholipids and lipid-like copolymers.

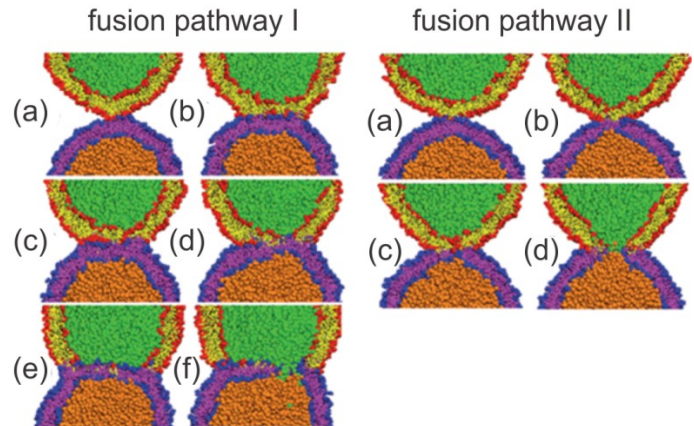

Figure 4. Fusion pathway I (left) for the low tension regime follows original stalk model with the fusion sequence: (a) kissing contact, (b) adhesion, (c) single-bilayer pore formation near the edge of the contact zone, (d) single-bilayer rupture, (e) hemifusion, and (f) fusion pore opening. Fusion pathway II (right) for the high tension regime belongs to a modified stalk model with the fusion sequence: (a) kissing contact, (b) stalk formation, (c) transmembrane fusion pore formation, and (d) fusion pore opening. Upper vesicle has red lipid heads, yellow tails, and green lumenal contents, while the lower one has blue lipid heads, purple tails, and gold lumenal contents. Simulation snapshots are from Ref. [97].

**2.3 RBC membrane models**
The aforementioned CG particle membrane models only simulate the lipid bilayer without considering the membrane cytoskeleton and thus limit their applications in the study of the biological problems regarding the RBC membrane, such as interactions between lipid bilayer and cytoskeleton, and between the cytoskeleton and diffusing membrane proteins. To model membrane for RBCs, Li et al. introduced the a two-component RBC membrane model by combining a cytoskeleton model with lipid bilayer model [102]. The cytoskeleton model is comprised of particles that represent actin junctional complexes, which are connected by the WLC potentials to represent spectrin filaments. Although this two-component model exhibits membrane shear modulus that is comparable with experimental measurements, the implicit representation of the spectrin filaments by WLC potential did not allow it to consider the interactions between lipid bilayer and cytoskeleton. Therefore, Li et al.[103] extended this two-component RBC membrane model by representing spectrin filaments explicitly by CG particles [104]. This modified membrane model describes the RBC membrane as a two-component system, including the cytoskeleton and the lipid bilayer. The lipid bilayer is represented by three types of CG particles and the cytoskeleton is represented by two types of CG particles

(see Fig. 5). The cytoskeleton consists of the hexagonal spectrin network and actin junctions. The actin junctions represented by the black particles (see Fig.5), are connected to the lipid bilayer via glycophorin (yellow particles). 39 spectrin particles (grey particles) connected by unbreakable springs are used to represent spectrin filament is represented by. The spring potential, $u_{cy}^{s-s}(r) = k_0(r - r_{eq}^{s-s})^2/2$, with equilibrium distance $r_{eq}^{s-s} = L_{max}/39$, where $L_{max}$ is the spectrin contour length and $r_{eq}^{s-s} \cong 5$ nm. The spectrin chain is connected to the band-3 particles (white particles) in the middle and are linked to the actin junction via spring potential, $u_{cy}^{a-s}(r) = k_0(r - r_{eq}^{a-s})^2/2$ at two ends. The equilibrium distance $r_{eq}^{a-s} = 10$ nm. The spring constant $k_0 = 57\ \varepsilon/\sigma^2$. A L-J potential is applied between spectrin particles,

$$u_{rep}(r_{ij}) = \begin{cases} 4\varepsilon\left[\left(\dfrac{\sigma}{r_{ij}}\right)^{12} - \left(\dfrac{\sigma}{r_{ij}}\right)^{6}\right] + \varepsilon & r_{ij} < R_{cut,LJ} \\ 0 & r_{ij} > R_{cut,LJ} \end{cases} \quad (1)$$

where $r_{ij}$ is the interparticle distance. Energy unit and the length unit are $\varepsilon$ and $\sigma$, respctively. The cutoff distance $R_{cut,LJ}$ is selected to be the equilibrium distances $r_{eq}^{s-s}$ between the spectrin particles.

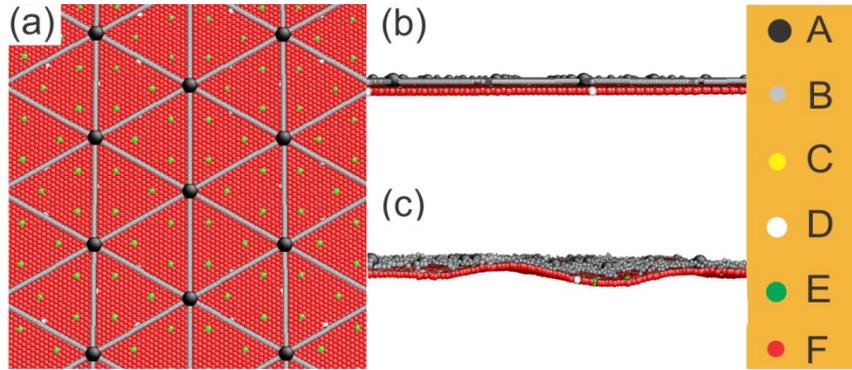

Figure 5. (a) top view and (b-c) side views of two-component RBC membrane model. "A" type particles represent actin junctions. "B" type particles represent spectrin proteins. "C" type particles represent glycophorin proteins. "D" type particles represent band-3 complex that are connected to the spectrin network. "E" type particles represent band-3 complex that are not connected to the network. "F" type particles represent lipid particles.

Lipid bilayer and transmembrane proteins are represented by three types of CG particles are introduced to represent (see Fig. 5). A cluster of lipid molecules is grouped into a red particle, which have a diameter of 5 nm. The yellow particles denote glycophorin proteins. One third of band 3 particles (white particles), denoting band 3 dimmers, are linked to the cytoskeleton. Two thirds of the band 3 particles (green particles) simulate the mobile band 3 proteins. Both translational and rotational degrees of freedom ($\mathbf{x}_i$, $\mathbf{n}_i$) are assigned to the CG particles, which form the lipid bilayer and transmembrane proteins. $\mathbf{n}_i$ and $\mathbf{x}_i$ are the direction and position vectors of particle $i$, respectively. $\mathbf{x}_{ij} = \mathbf{x}_j - \mathbf{x}_i$ is defined as the interparticle distance vector between $i$ and $j$. Correspondingly, $r_{ij} \equiv |\mathbf{x}_{ij}|$ is the distance, and $\hat{\mathbf{x}}_{ij} = \mathbf{x}_{ij}/r_{ij}$ is a unit vector. The CG particles, forming the lipid membrane and membrane proteins, interact through a pair-wise soft potential

$$u_{mem}(\mathbf{n}_i, \mathbf{n}_j, \mathbf{x}_{ij}) = u_R(r_{ij}) + A(\alpha, a(\mathbf{n}_i, \mathbf{n}_j, \mathbf{x}_{ij})) u_A(r_{ij}), \tag{2}$$

$$\begin{cases} u_R(r_{ij}) = k\varepsilon \left((R_{cut,mem} - r_{ij})/(R_{cut,mem} - r_{eq})\right)^8 - k\varepsilon & \text{for } r_{ij} < R_{cut,mem} \\ u_A(r_{ij}) = -2k\varepsilon \left((R_{cut,mem} - r_{ij})/(R_{cut,mem} - r_{eq})\right)^4 - k\varepsilon & \text{for } r_{ij} < R_{cut,mem} \\ u_R(r_{ij}) = u_A(r_{ij}) = 0, & \text{for } r_{ij} \geq R_{cut,mem} \end{cases} \tag{3}$$

where $u_A(r_{ij})$ and $u_R(r_{ij})$ are the attractive and repulsive parts of the pair potential, respectively. $\alpha$ is a parameter that can tune the attractive part of the potential. Fig. 6(a) illustrates that when $A(\alpha,a) = +1$, the energy well of the applied potential is wider than the LJ6-12 potential, thsu facilitating CG particles pasting each other and thus enhance the fluidity of membrane. The function $A(\alpha,a(\mathbf{n}_i,\mathbf{n}_j,\mathbf{x}_{ij})) = 1+\alpha(a(\mathbf{n}_i,\mathbf{n}_j,\mathbf{x}_{ij}) -1)$ adjusts the energy well of the potential through which regulating the fluidity of the membrane, as shown in Fig.6(b). In the simulations, $\alpha$ was chosen to be 1.55 and cutoff distance $R_{cut,mem}$ is $2.6\sigma$. The parameters $\alpha$ and $R_{cut,mem}$ are selected to maintain the fluid phase of the lipid bilayer. The directional vector carried by each CG particle is important for the self-assembly of the membrane in the absence of solvent particles and lipid chains. The detailed information about applied potentials and the selection of the potential parameters can be found from authors' previous work [102, 103].

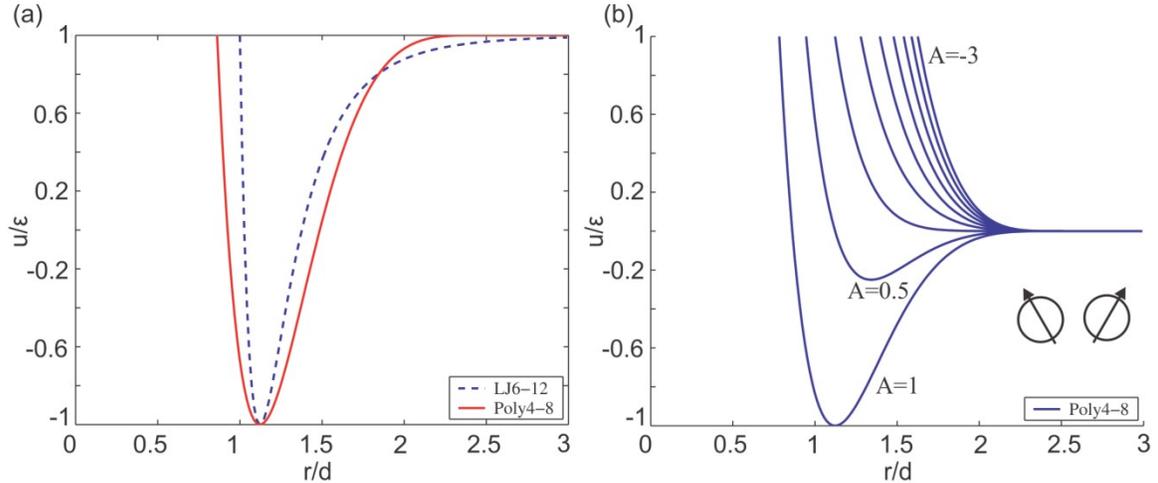

Figure 6. The pair-wise interaction potential expressed by Eq (3). (a) Comparison between the LJ6-12 potential and applied potential; (b) the interaction potential plotted as a function of $A(\alpha,a)$. Adapted from [102, 105].

This modified two-component membrane model can simulate the molecular structures of the normal or diseased RBC membrane and thus has been applied to study membrane stiffness, the protein diffusion and membrane vesiculation (see Fig.7) in the healthy and pathological conditions [105-108].

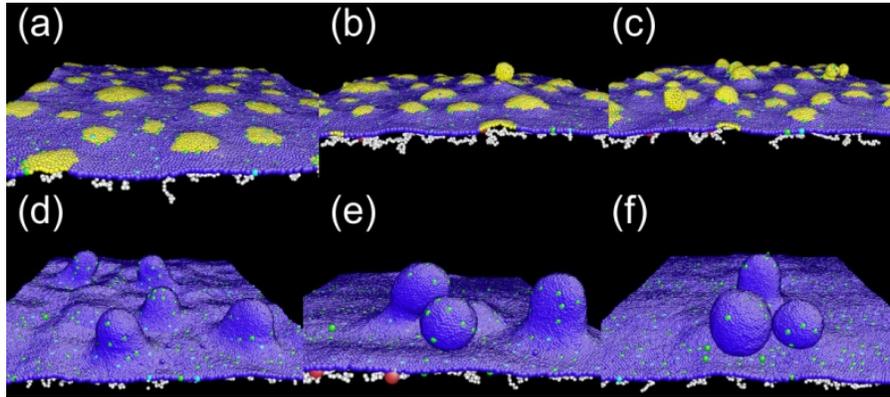

Figure 7. (a-c) Membrane budding and nanovesiculation induced by membrane domain spontaneous curvatures. Domains are plotted by the yellow particles). (a) The membrane domains only bulge out from the membrane but no vesicle is formed when the spontaneous curvature is small. (b) Vesiculation occurs as spontaneous curvature is increased. (c) When spontaneous curvature is further increased, more vesicles are observed. (d-f) Vesiculation induced by membrane compression (d) At the beginning of the compression, only one protuberance is created at the beginning of the compression. (e) The protrusion grows and a vesicle is formed as the compression continues. Subsequently, a new bud is generated and then forms a vesicle. (f) Eventually, two vesicles are obtained from compressing the membrane. Adapted from [105]

However, it was computationally expensive to simulate large membrane domains such as an entire RBC or whole blood involving large numbers of RBCs. To increase the length scale of the modified two-component membrane model, Tang et al. [109] recently developed OpenRBC, a multi-thread CGMD code, which is capable of simulating an entire RBC with explicit representations of lipid bilayer and cytoskeleton by multiple millions of CG particles using a single shared memory commodity workstation, as shown in Fig. 8.

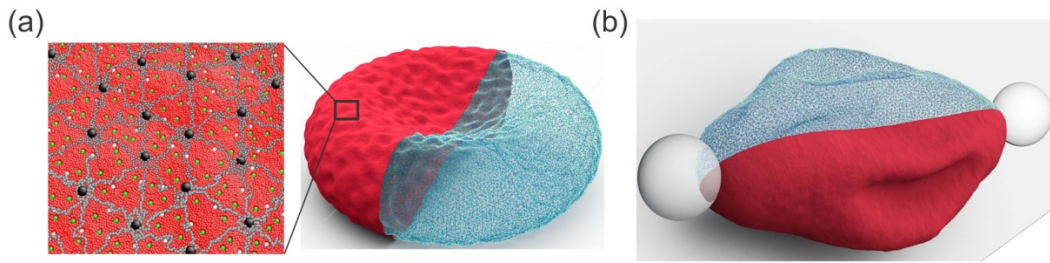

Figure 8. (a) OpenRBC can simulate the entire RBC based on the two-component RBC membrane model with explicit representations of lipid bilayer and cytoskeleton. Adapted from [109] (b) Simulation of optical tweezer experiment on a RBC by using OpenRBC. Adapted from [109].

**2.4 RBC whole cell models**

Although OpenRBC allows us to simulate an entire RBC at protein resolution, it is still computationally prohibitively to simulate RBC suspensions and study the blood rheology. Thus, a more efficient RBC model with higher level of coarse-graining is required. Discher and co-workers [110] developed a CG particle model for the RBC membrane's spectrin cytoskeleton to the study the elasticity of RBC in micropipette aspiration. In this spectrin-level RBC model, a RBC is treated as a 2D canonical hexagonal network of CG particles (see Fig.9(a)). The neighboring CG particles are connected via a WLC potential to represent a spectrin filament. The bending rigidity induced by the lipid bilayer is represented by a bending potential applied between two neighboring triangles. In addition, constrains on cell volume and surface area are implemented to model the incompressibility of the cytosol and lipid bilayer. Li et al. [111] extended this model by introducing a dynamic spectrin network, spontaneous curvature of the lipid membrane and in-plane shear energy relaxation. Pivkin et al. [112] introduced hydrodynamic effect into this elastic RBC model by combining it with DPD. More importantly, this RBC model is further coarse-grained from 23867 CG particles to 500 CG particles while the elastic properties are still preserved. This DPD-based whole cell model is computationally very efficient and it can be used not only in simulations of the single cell dynamics [113] and morphological changes of diseased RBC (see Fig.9(b)) [114], but also the rheology of the blood flow [115, 116] and blood flowing through micro-devices (see Fig.9(c))[117]. However, it is not capable of modeling the interactions between the cytoskeleton and the lipid bilayer, which is frequently essential, as demonstrated in the study of the lipid bilayer detachment and RBC vesiculation. To overcome this issue, a RBC model made of two layers of 2D triangulated networks developed to study lipid bilayer and cytoskeleton interactions in healthy and diseased RBCs[118-120]. Both one-component and two-component whole-cell models consider the bending energy ($V_b$), the elastic energy ($V_s$), constraints of enclosed volume ($V_v$) and fixed surface area ($V_a$) of an RBC. In the two-component whole-cell model, bilayer-cytoskeleton interaction energy ($V_{int}$) is additionally considered in two-component models. $V_s$ represents the elastic spectrin network, given by

$$V_s = \sum_{j \in 1 \ldots N_s} \left[ \frac{k_B T l_m \left(3x_j^2 - 2x_j^3\right)}{4p(1-x_j)} + \frac{k_p}{(n-1)l_j^{n-1}} \right] \qquad (4)$$

where $l_m$ is the maximum spring extension while $l_j$ is the length of the spring $j$. $x_j = l_j/l_m$. $p$ represents the persistence length. $k_B T$ designates the energy unit. $k_p$ is the spring constant of the potential. $n$ is a specified exponent. $V_b$ results from the bending stiffness of lipid bilayer and it is expressed as

$$V_b = \sum_{j \in 1...N_s} k_b[1 - cos(\theta_j - \theta_0)] \qquad (5)$$

where $k_b$ is the bending constant. $\theta_j$ is the instantaneous angle between two neighboring triangles. $\theta_0$ is the spontaneous angle between the two neighboring triangles. $V_a$ and $V_v$ are implemented to represent the incompressible lipid bilayer and interior fluid. The potentials for constraining the surface area and volume are given by

$$V_a = \sum_{j \in 1...N_t} \frac{k_d(A_j - A_0)^2}{2A_0} + \frac{k_a(A_{cell} - A_{cell,0}^{tot})^2}{2A_{cell,0}^{tot}} \qquad (6)$$

$$V_v = \frac{k_v(V_{cell} - V_{cell,0}^{tot})^2}{2V_{cell,0}^{tot}} \qquad (7)$$

where $A_0$ is the triangle area. $N_t$ is the number of triangles in the membrane network. $k_d$, $k_a$ and $k_v$ are the constraint coefficients for the local area, global area and volume, respectively. The specified total area and volume are $A_{cell,0}^{tot}$ and $V_{cell,0}^{tot}$, respectively. $V_{int}$ represents the interaction between the lipid bilayer and it is expressed as a summation of harmonic spring potentials

$$V_{int} = \sum_{j,j' \in 1...N_{bs}} \frac{k_{bs}(d_{jj'} - d_{jj',0})^2}{2} \qquad (8)$$

where $N_{bs}$ is the number of bond connections between the lipid bilayer and the cytoskeleton. $k_{bs}$ is the spring constant. $d_{jj'}$ measures the distance between the vertex $j$ on the cytoskeleton and its projection point $j'$ on the lipid bilayer. The vector $n_{jj'}$ and $d_{jj',0}$ is the initial distance between the vertex $j$ and the point $j'$.

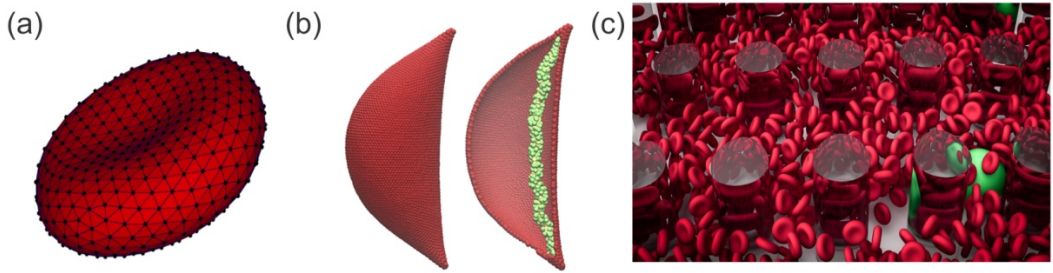

Figure 9. (a) A DPD whole cell model where RBC the membrane is modeled as a 2D canonical hexagonal network of DPD particles and the neighboring DPD partciles are connected via a WLC potential that represents a spectrin filament. (b) Application DPD whole cell model on simulation the cell morphology change due to the presence of sickle hemoglobin fibers in sickle cell disease. Adapted from [114]. (c) DPD whole cell model was applied to simulate RBCs passing pillar matrix on a mcirodevice. Adapted from [117].

## 5. Summary

Although atomistic simulations provide detailed structural properties of lipid membranes, they are limited to studying small membrane patches owing to high computational expense. Many interesting biological phenomena can not be accessed by these simulations when individual atoms or lipid molecules are resolved. Continuum-based methods can model bigger membrane or entire RBCs. However, they can not address problems concerning the detailed structural changes in the RBC membrane and specific protein defects in diseased RBCs. In addition, typical continuum models are not suitable to simulate dramatic topology changes such as membrane budding, fusion, and self-assembly of lipids to vesicles. CG particle models are bridging in certain aspects the MD and continuum methods. First, it is a discrete CG particle-based method similar to MD, but with each CG particle representing a lump of molecules, thus simplifying the atomistic dynamics by eliminating fast degrees of freedom while preserving the mesoscopic structures and properties of membrane or RBCs, via which it provides an efficient simulation means to capture correct dynamics of membrane or RBCs at larger spatial and temporal scales. Second, the CG particles can be treated as a mesh-free representation of continuum. Particularly, CG particle model can be used to simulate dynamical processes involve significant topological change, such as lipid particles self-assembly, vesicle fusion and nanoparticle transport through the erythrocyte membrane. It is worth noting that in CGMD simulations, the employed length scale and time scale do not have an immediate correlation with a physical system. Such a correspondence can be established only through comparison with a natural procedure.. An emerging particle-based method, smoothed dissipative particle dynamics (SDPD) [121-123], has also been applied to modeling rheology of red and white blood cell behavior to elaborate the interaction of the surrounding fluid and membrane [124, 125]. SDPD method has advantages over conventional DPD, including well-defined physical scale of particles, direct inputs of transport properties and arbitrary equation of state. As an alternative of DPD, SDPD can be applied to study the transport phenomenon involved in the blood flows, such as drug delivering and cell loading in saline solution. Computational simulations have become increasingly important to improve our understanding of the rheology and dynamics of RBCs, in particular in blood diseases. The CG membrane and RBC models developed in the last decade have been successfully implemented to uncover various membrane functions and characteristics. A complete picture of applications of CG particle models for membrane and RBC in physiological and pathological conditions and relevant references can be found in recent reviews [126-128].

**Acknowledgments**

The work described in this article was supported by National Institutes of Health grant No. U01HL114476.